\documentclass{an}
\usepackage{graphicx}
\usepackage{times}
\usepackage{fancyhdr}
\usepackage{natbib}
\newcommand{\quot}[1]{`#1'}

\newcommand{\D}[1]{D_{#1}}
\newcommand{\A}{\widehat{A}}

\newcommand{\obsom}{\omega^\star}
\newcommand{\imp}[1]{\textit{#1}}
\newcommand{\const}{\mathrm{const}}

\bibpunct{(}{)}{;}{}{}{,}

\begin{document}

\title{The autoparametric 3:2 resonance in conservative systems}

\author{Ji\v{r}\'{\i} Hor\'ak}
\institute{Astronomical Institute, The Czech Academy of Sciences, 
    Bo\v{c}n\'{\i}~II, CZ-140\,31~Prague, Czech Republic}

\date{Received; accepted; published online}

\abstract{In the resonance model, high-frequency quasi-periodic oscillations 
(QPOs) are supposed to be a consequence of nonlinear resonance between
modes of oscillations occurring within the innermost parts of an
accretion disk.  Several models with a prescribed mode--mode interaction
were proposed  in order to explain the characteristic properties of the 
resonance in QPO sources. In this paper, we examine nonlinear
oscillations  of a system having two degrees of freedom and we show that
this case could be particularly relevant for QPOs. We present a very
convenient way how to study  autoparametric resonances of a fully
general system using the method of multiple  scales. We concentrate to
conservative systems and discuss their behavior near the 3:2
parametric resonance.
}

\correspondence{horak@astro.cas.cz}

\maketitle

\section{Introduction}
In the resonance model \citep{ak01,ka01,kato03}, there is a natural and
attractive possibility of explaining the observed rational ratios of
high-frequency QPOs as a consequence of non-linear coupling  between
different modes of accretion disk oscillations. The idea has been
pursued in several papers \citep[recently, e.g.][]{akklr03, r04}. 

Specific models invoke particular physical mechanisms. Some models can
be almost immediately comprehended as distinct realizations of the
general approach discussed here -- for example, various formulations of
the orbiting spot model \citep{sb04} or the models, where QPOs are
produced by the magnetically driven resonance  in a diamagnetic
accretion disk \citep{lai99} -- while other seem to be more distant from
the view presented herein -- e.g.\ the transition layer model
\citep{tit02}, an interesting idea of p-mode oscillations of a small
accretion torus \citep{rezzola03} or the model of blobs in an accretion
disc \citep[see e.g.][and references cited therein]{ka99,li04}.  Also in
this context, \citet{kato04} discussed the resonant interaction  between
waves propagating in a warped disk, including their rigorous
mathematical description. Instead of pursuing a specific model, here we
keep the discussion as general as possible, aiming to implement the
formalism of multiple scales. Indeed, we show that there is
unquestionable appeal in this approach which offers some additional
insight into generic properties of resonant oscillations.

Some properties of an accretion disk oscillations can be  discussed
within the epicyclic approximation of a test particle on a circular
orbit near equatorial plane.  Suppose that angular momentum of the
particle is fixed to a value $ \ell $. The effective potential $
U_\ell(r, \theta) $ has a minimum at radius $ r_0 $, corresponding to
the location of the stable circular orbit. An observer moving along this
orbit measures radial, vertical and azimuthal epicyclic oscillations of
a particle nearby. Since the angular momentum of the particle is
conserved, only two of them -- radial and vertical -- are independent.
The epicyclic frequencies can be derived from the geodesic equations
expanded to the linear order in deviations $ \delta r = r - r_0 $ and $
\delta \theta = \theta - \pi/2 $ from the circular orbit. We get two
independent second-order differential equations describing two uncoupled
oscillators with frequencies $ \omega_r $ and $
\omega_\theta $, which are given by the second derivatives of effective
potential $ U_\ell(r, \theta) $. 

In Newtonian theory, $ \omega_r $ and $ \omega_\theta $ are equal to the
Keplerian orbital frequency $ \Omega_K $. This is in tune with the fact
that orbits of particles are planar and closed curves. The degeneracy
between two epicyclic frequencies can be seen as a result of
scale-freedom of the Newtonian gravitational potential \citep{ak03}. In
Schwarzschild geometry this freedom is broken by introducing the
gravitational radius $ r_g = 2GM/c^2 $. The degeneracy between the
vertical epicyclic and the orbital frequencies is related to spherical
symmetry of the gravitational potential, which assures the existence of
planar trajectories of particles. All three frequencies are different in
the vicinity of a rotating Kerr black hole.

In addition, when nonlinear terms of geodesic equations are included,
the  two oscillations in $r$ and $\theta$ directions become coupled and 
variety of new phenomena connected to nonlinear nature of the equations 
appear. This rich phenomenology includes frequency shift of observed
frequencies  with respect to eigenfrequencies, presence of higher
harmonics and  subharmonics, drifts and parametric resonance. The first 
three are connected to nonlinear oscillations of each mode and the last
one  comes from the coupling between two modes.

\section{Expansion via multiple scales}
\label{sec:res}

We study nonlinear oscillations of the system having two degrees of
freedom, i.e., the coordinate perturbations $\delta r$ and  $\delta
\theta$. The oscillations are described by two coupled differential
equations of the very general form
\begin{eqnarray}
  \label{eq:res_gov_r}
  \ddot{\delta r} + \omega_r^2\; \delta r &=& \omega_r^2\; f_r(\delta r,
  \delta\theta, \dot{\delta r},\dot{\delta \theta}), 
  \\
  \label{eq:res_gov_theta}
  \ddot{\delta \theta} + \omega_\theta^2\; \delta \theta &=&
  \omega_\theta^2
  \;f_\theta(\delta r, \delta\theta, \dot{\delta r}, \dot{\delta \theta}).
\end{eqnarray}
Suppose that the functions $ f_r $ and $ f_\theta $ are nonlinear, i.e.,
their Taylor expansions start in the second order. Our another
assumption is that these functions are invariant under reflection of
time (i.e., the Taylor expansion does not contain odd powers of time
derivatives of $ \delta r $ and $ \delta \theta $). As we shell see
later, this  is related to the conservation of the total energy in the
system. Many authors studied such systems with a particular form of
functions $ f $ and $ g $ \citep{nm79}, however, in this paper we keep
the discussion fully general. 

We seek the solutions of the governing equations in the form of the
multiple-scales expansions \citep{nm79}
\begin{equation}
  \delta r(t, \epsilon) = \sum_{n=1}^4 \epsilon^n r_n(T_\mu),
  \quad
  \delta \theta(t, \epsilon) = \sum_{n=1}^4 \epsilon^n \theta_n(T_\mu),
  \label{eq:res_exp}
\end{equation}
where several time scales $T_\mu$ are introduced instead of the physical
time $t$,
\begin{eqnarray}
  T_\mu \equiv \epsilon^\mu t, &  \mu = 0,1,2,3.
  \label{eq:ms_scales}
\end{eqnarray}
The time scales are treated as independent. It follows that instead of
the single time derivative we have an expansion of partial derivatives
with respect to the $ T_\mu $
\begin{eqnarray}
  \frac{d}{dt} &=& \D{0} + \epsilon \D{1} + \epsilon^2 \D{2} +
  \epsilon^3 \D{3} +  {\cal O}(\epsilon^4), 
  \label{eq:ms_der1}
  \\
  \frac{d^2}{dt^2} &=& \D{0}^2 + 2 \epsilon \D{0} \D{1} + 
  \epsilon^2 (\D{1}^2 + 2 \D{0}\D{2}) + \
  \nonumber \\
  &\phantom{=}& 2\epsilon^3 (\D{0}\D{3} + \D{1}\D{2}) + {\cal
  O}(\epsilon^4),
  \label{eq:ms_der2}
\end{eqnarray}
where $ \D{\mu} = \partial / \partial T_\mu $.

We expand the nonlinear functions $f_r$ and $f_\theta$ into the Taylor 
series and then we substitute the expansions (\ref{eq:res_exp}),
(\ref{eq:ms_der1}) and (\ref{eq:ms_der2}). Finally, we compare the
coefficients of the same powers of $\epsilon$  on both sides in the
resulting couple of equations. This way we get a set of \imp{linear}
second-order differential equations that can be solved successively --
the lower-order  terms of the expansion (\ref{eq:res_exp}) appear as
forcing terms on the right-hand sides of the equations for the higher
order approximations. 

In the first order we obtain equations corresponding to the linear
approximation
\begin{equation}
  \label{eq:res_1}
  (\D{0}^2 + \omega_r^2) r_1 = 0,
  \quad
  (\D{0}^2 + \omega_\theta^2) \theta_1 = 0.
\end{equation}
with the solutions
\begin{eqnarray}
  \label{eq:res_1solr}
  x_1 = A_r(T_1,T_2,T_3) e^{i \omega_r T_0} + \mathrm{cc},
  \\
  \theta_1 = A_\theta(T_1,T_2,T_3) e^{i \omega_\theta T_0} + \mathrm{cc}.
  \label{eq:res_1solth}
\end{eqnarray}
The complex amplitudes $\A_r$ and $A_\theta$ generally depend on the
higher time-scales. 

The solutions (\ref{eq:res_1solr}) and (\ref{eq:res_1solth}) substituted
into the quadratic terms on the right-hand sides of the second-order
differential equations produce terms that oscillates  with frequencies
$2\omega_r$, $2\omega_\theta$ and $\omega_\theta\pm\omega_r$. When the
frequency  ratio $\omega_r/\omega_\theta$ is far from 1:2 and 2:1 the
particular solutions $r_2$ and $\theta_2$  describe higher harmonics to
the linear-order oscillations $r_1$ and $\theta_1$. Hence, the presence
of  higher harmonics  in the power-spectra is a signature of nonlinear
oscillations. Their frequencies and relative strengths with respect to
the main oscillations provide us an usefull informations about 
nonlinearities in the system. 

In addition, the right hand sides of the second order equations contain
terms proportional to $e^{i\omega_r T_0}$ and $e^{i\omega_\theta T_0}$
that oscillates with the same frequency as the eigenfrequency of the
oscillators. These terms produces secular grow of the amplitudes of the 
second-order approximations $r_2$ and $\theta_2$ and causes nonuniform
expansions (\ref{eq:res_exp}). Eliminating them we get the
\imp{solvability conditions} for the complex amplitudes
$A_r(T_1,T_2,T_3)$ and $A_\theta(T_1,T_2,T_3)$ that give us the
evolution of the system on longer time-scales \citep{nm79}.

When the eigenfrequencies are in 1:2 or 2:1 ratio we observe
qualitatively different behavior related  to the \imp{autoparametric
resonance}. In that case the right hand sides contains additional
secular terms and the solvability conditions take different form.
Different resonances occur in different orders of approximation. The
possible resonances in the third order are 1:3, 1:1 and 3:1 and 1:4,
3:2, 2:3 and 4:1 in  the fourth order\footnote{The ratio $n:m$ refers to
the eigenfrequency ratio $\omega_\theta:\omega_r$.} However, if the
governing equations remain unchanged under the transformation 
$\delta\theta\rightarrow-\delta\theta$ (i.e., the system is reflection
symmetric) the only autoparametric resonances that exists in the system
are 1:2, 1:1, 1:4 and 3:2 \citep{r04}

\section{The 3:2 autoparametric resonance}
\label{sec:32}

Let us consider oscillations of a conservative system eigenfrequencies
of which are close to 3:2.  The time behavior of the observed
frequencies $\obsom_r$ and $\obsom_\theta$ and amplitudes $a_r$  and
$a_\theta$ of the oscillations can be found from the solvability
conditions imposed on the  complex amplitudes $A_r(T_1,T_2,T_3)$ and
$A_\theta(T_1,T_2,T_3)$ \citep{hor05}
\begin{eqnarray}
  \D{1}A_r &=& \D{2}A_\theta = 0,
  \label{eq:sol1}
  \\
  \D{2} A_r &=& - \frac{i \omega_r}{2} \left[ \kappa_r |A_r|^2 +
  \kappa_\theta |A_\theta|^2 \right] A_r,
  \\
  \D{2} A_\theta &=& - \frac{i \omega_\theta}{2} \left[ \lambda_r
  |A_r|^2 +
  \lambda_\theta |A_\theta|^2 \right] A_\theta,
  \\
  \D{3} A_r &=& -\frac{i}{2}\omega_r \alpha (A_r^2)^\ast A_\theta^2
  e^{-i (\sigma_2 T_2 + \sigma_3 T_3)},
  \\
  \D{3} A_\theta &=&-\frac{i}{2}\omega_\theta \beta A_r^3 A_\theta^\ast 
  e^{i(\sigma_2 T_2 + \sigma_3 T_3)}.
  \label{eq:sol5}  
\end{eqnarray}
In the fourth order we eliminate also terms which become secular in when
$3\omega_r\approx2\omega_\theta$.  We describe vicinity of the resonance
by the detuning parameters $\sigma_2$ and $\sigma_3$ introduced
according to
\begin{equation}
  3 \omega_r = 2 \omega_\theta + \epsilon^2 \sigma_2 + \epsilon^3
  \sigma_3.
\end{equation}
The term $ \epsilon \sigma_1 $ is missing, because the complex
amplitudes depends only on the second and  the third time-scales. The
solvability conditions describe evolution of the system in the most
general way: the real parameters $\alpha$, $\beta$, $\kappa_r$,
$\kappa_\theta$, $\lambda_r$ and $\lambda_\theta$ are given  by the
coefficients of the Taylor-expanded nonlinear functions $f_r$ and
$f_\theta$. 

Since $A_r$ and $A_\theta$ are complex, the conditions
(\ref{eq:sol1})--(\ref{eq:sol5}) represents 12 real equations. However
few of them are trivial. By substituting the polar forms  $\epsilon
A_r=\frac{1}{2}a_r e^{i \phi_r}$ and $\epsilon
A_\theta=\frac{1}{2}a_\theta e^{i \phi_\theta}$,  separating real and
imaginary parts and introducing the unique time $t$ the number of the
equations can be reduced to four,
\begin{eqnarray}
  \dot{a}_\rho &=& 
  \frac{\alpha\omega_r}{16}\, a_\rho^2\,a_\theta^2\,\sin \gamma, 
  \label{eq:qpo-ar} \\
  \dot{a}_\theta &=& 
  -\frac{\beta \omega_\theta}{16}\,a_\rho^3\,a_\theta\,\sin \gamma, 
  \label{eq:qpo-atheta} \\
  \dot{\phi}_\rho &=&
  -\frac{\omega_r}{2}\, 
  \left[\kappa_r\,a_\rho^2 + \kappa_\theta\,a_\theta^2 \right] -
  \frac{\alpha \omega_r}{16}\, a_\rho\,a_\theta^2\,\cos \gamma,
  \label{eq:qpo-phir} \\
  \dot{\phi}_\theta &=& 
  -\frac{\omega_\theta}{2}\, \left[\lambda_r\,a_\rho^2 +
  \lambda_\theta\, a_\theta^2 \right] - \frac{\beta \omega_\theta}{16}
  a_\rho^3\,\cos \gamma,
  \label{eq:qpo-phitheta}
\end{eqnarray}
where we introduced the phase function
$\gamma(t)=2\phi_\theta(t)-3\phi_r(t)-\sigma t$ and the  unique detuning
parameter $\sigma=\epsilon^2\sigma_2+\epsilon^3\sigma_3$.  The equations
(\ref{eq:qpo-ar}) and (\ref{eq:qpo-atheta}) describes the slow evolution
of the amplitudes of oscillations and additional long-term behavior of
the oscillation phases is given by equations (\ref{eq:qpo-phir})  and
(\ref{eq:qpo-phitheta}). These equations give us the frequency-shift of
the observed frequencies $\obsom_r$ and $\obsom_\theta$ with respect to
the eigenfrequencies $\omega_r$ and $\omega_\theta$, respectively,
\begin{eqnarray}
  \omega^\ast_r = \omega_r + \dot{\phi}_r, \quad
  \omega^\ast_\theta = \omega_\theta + \dot{\phi}_\theta.
  \label{eq:qpos-corrections}
\end{eqnarray}

The two equations (\ref{eq:qpo-phir}) and (\ref{eq:qpo-phitheta}) can be
replaced by a single differential equation for the phase function,
\begin{equation}
  \dot{\gamma}=-\sigma+\frac{\omega_\theta}{4}\left[\mu_r a_r^2 +
  \mu_\theta a_\theta^2 + \frac{a_r}{2} \left( \alpha a_\theta^2 - \beta
  a_r^2 \right) \cos \gamma \right],
  \label{eq:qpo-gamma}
\end{equation}
were we used the fact that near the resonance $ \omega_r \approx (2/3)
\omega_\theta $ and we defined $ \mu_r = \kappa_r - \lambda_r $ and
$ \mu_\theta = \kappa_\theta - \lambda_\theta $.

\subsection{Steady-state solutions}

Steady-state solutions are characterized by constant amplitudes and
frequencies of oscillations. Such solutions represent singular points of
the system governed by equations (\ref{eq:qpo-ar}),
(\ref{eq:qpo-atheta}) and (\ref{eq:qpo-gamma}).

 It is obvious from equations (\ref{eq:qpo-ar}) and
(\ref{eq:qpo-atheta}) that the condition  $\dot{a}_r= \dot{a}_\theta=0 $
can be satisfied (with nonzero amplitudes) only if $\sin \gamma = 0$
(identically at all times), and thus also $\dot{\gamma}=0 $. In that
case equation (\ref{eq:qpo-gamma})  transforms to the algebraic equation
\begin{equation}
  \frac{\sigma}{\omega_\theta} = \frac{1}{4} \left[\mu_r a_r^2 +
  \mu_\theta a_\theta^2 \pm \frac{a_r}{2} \left( \alpha a_\theta^2 - \beta
  a_r^2 \right)\right].
\end{equation}
The left-hand side can be expressed using the eigenfrequency ratio
$R=\omega_\theta/\omega_r$ as
\begin{equation}
  \frac{\sigma}{\omega_\theta} = -\frac{2}{R} \left( R - \frac{3}{2}
  \right).
\end{equation}
Then we get
\begin{equation}
  R = \frac{3}{2} - \frac{3}{16} \left( \mu_r a_r^2 + \mu_\theta
  a_\theta^2 \right) \pm \frac{3}{32} a_r \left( \alpha a_\theta^2 - \beta
  a_r^2 \right),
\end{equation}
were we neglected terms of the fourth order. Note that the lowest
correction to eigenfrequencies is of the second order -- for given
amplitudes $a_r$, $a_\theta$ the steady-state oscillations occur when
the eigenfrequency ratio departs from $3/2$ by deviation of order of
$a^2$.

The relation between observed frequencies of oscillations $\obsom_r$,
$\obsom_\theta$ and eigenfrequencies $\omega_r$, $\omega_\theta$
are given by the time derivative of phases $\phi_r$ and $\phi_\theta$.
We can find simple relation between observed frequencies and the
phase function
\begin{eqnarray}
  3\obsom_r - 2\obsom_\theta &=& 3\omega_r - 2\omega_\theta + (3\dot{\phi}_r
  - 2\dot{\phi}_\theta)
  \nonumber\\
  &=& \sigma + (3\dot{\phi}_r - 2\dot{\phi}_\theta) =
  -\dot{\gamma}.
  \label{eq:relation}
\end{eqnarray}
For steady state solutions $\dot{\gamma}=0$, and thus observed
frequencies are adjusted to \imp{exact} 3:2 ratio even if
eigenfrequencies depart from it. 

\subsection{Integrals of motion}

\begin{figure}
\includegraphics[width=0.5\textwidth]{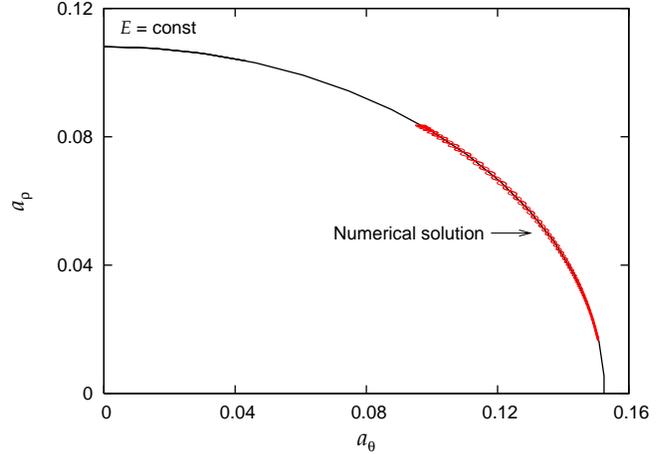}
\caption{
  Comparison between an analytical constraint (\ref{eq:32_E})  and the
  corresponding numerical solution of the system \citet{akklr03}.  Each
  point corresponds to the amplitudes of the oscillations at a
  particular time. On the other hand, from the discussion of equation
  (\ref{eq:32_E}) we know that these points must lay on an ellipse,
  whose shape is determined by the multiple-scales method.}
\label{fig:32_ellipse}
\end{figure}

The method of investigation of time-dependent behavior of the system is
analogical to the case of 1:2 resonance as examined by \citet{nm79}. The
oscillations are described by three variables $ a_r(t) $, $a_\theta(t)$
and $\gamma(t)$  and three first-order differential equations
(\ref{eq:qpo-ar}), (\ref{eq:qpo-atheta}) and (\ref{eq:qpo-gamma}).
However, the number of differential equations can be reduced to one
because it is possible two find two integrals of motion.  Our discussion
will be 

Consider equations (\ref{eq:qpo-ar}) and (\ref{eq:qpo-atheta}).
Eliminating $ \sin \gamma $ from both equations we find
\begin{equation}
  \frac{d}{dt}(a_r^2 + \nu a_\theta^2) = 0
\end{equation}
and thus
\begin{equation}
  \label{eq:32_E}
  a_r^2 + \nu a_\theta^2 = \mathrm{const} \equiv E,
\end{equation}
where we defined 
\begin{equation}
  \label{eq:32_nu}
  \nu = \frac{\alpha \omega_r}{\beta \omega_\theta} \approx \frac{2
  \alpha}{3 \beta}.
\end{equation}
When $ \nu > 0 $, the both amplitudes of oscillations are bounded. The
curve $ [a_r(t), a_\theta(t)] $ is a segment of an ellipse. The constant
$ E $ is proportional to the energy of the system. On the other hand,
when $ \nu < 0 $, one amplitude of oscillations can grow without bounds
while the second amplitude vanishes. This case corresponds to the
presence of an regenerative element in the system \citep{nm79}. The
corresponding curve in the $(a_r,a_\theta)$ plane is a hyperbola. In
further discussion we assume that $ \nu > 0 $. 

In order to verify that the the energy of the system is conserved, we
numerically integrated governing equation (\ref{eq:res_gov_r}) and
(\ref{eq:res_gov_theta}) for the one particular system discussed by
\citet{akklr03}. The comparison is in Figure \ref{fig:32_ellipse}. 
The numerical and analytical results are in a very good agreement.

The second integral of motion is found in following way. Let us multiply
equation (\ref{eq:qpo-gamma}) by $a_\theta$. Then we obtain
\begin{eqnarray}
  a_\theta \dot{\gamma} &=& -\sigma a_\theta  +  \frac{\omega_\theta}{4} 
  \mu_r a_r^2 a_\theta  +  \frac{\omega_\theta}{4} \mu_\theta a_\theta^3 
  +  \frac{\omega_\theta}{8} \alpha a_r a_\theta^3 \cos \gamma -
  \nonumber\\ 
  &\phantom{=}&\frac{\omega_\theta}{8} \beta a_r^3 a_\theta \cos \gamma.
\end{eqnarray}
Changing the independent variable from $t$ to $a_\theta$ and multiplying
the whole equation by $d a_\theta$ we find
\begin{eqnarray}
  \label{eq:32_derL1}
  a_r^3 a_\theta^2 d(\cos \gamma) +  \frac{8\sigma}{\beta \omega_\theta}
  d(a_\theta^2) \!\!&-&\!\!  \frac{4 \mu_r}{\beta} a_r^2 a_\theta
  d(a_\theta^2) -
  \frac{\mu_\theta}{\beta} d(a_\theta^4) - 
  \nonumber
  \\
  -\frac{2\alpha}{\beta} a_r a_\theta^3 \cos \gamma d a_\theta
  \!\!&+&\!\! 2 a_r^3 a_\theta \cos \gamma d a_\theta = 0.
\end{eqnarray}
The equation (\ref{eq:32_E}) implies 
\begin{equation}
  \label{eq:32_dE} a_\theta d a_\theta = -\frac{a_r d a_r}{\nu}. 
\end{equation}
With the aid of this relation equation (\ref{eq:32_derL1}) takes the
form
\begin{eqnarray}
  3 a_r^2 a_\theta^2 \cos \gamma d a_r  +  2 a_r^3 a_\theta \cos \gamma d
  a_\theta  +  a_r^3 a_\theta^2 d(\cos\gamma)  +   
  \nonumber
  \\
   + \frac{8\sigma}{\beta \omega_\theta} d(a_\theta^2) + 
  \frac{\mu_r}{\beta \nu} d (a_r^4)  -  \frac{\mu_\theta}{\beta}
  d(a_\theta^4)  = 0.
\end{eqnarray}
The first three terms express the total differential of function $-a_r^3
a_\theta^2 \cos \gamma $. Hence, the above equation can be arranged to
the form
\begin{equation}
  d \left( a_r^3 a_\theta^2 \cos \gamma  +  \frac{8\sigma}{\beta
  \omega_\theta} a_\theta^2  +  \frac{\mu_r}{\beta \nu} a_r^4  - 
  \frac{\mu_\theta}{\beta} a_\theta^4 \right) = 0.
\end{equation}
In other words,
\begin{equation}
  \label{eq:32_L}
  a_r^3 a_\theta^2 \cos \gamma  +  \frac{8\sigma}{\beta \omega_\theta}
  a_\theta^2  +  \frac{\mu_r}{\beta \nu} a_r^4  - 
  \frac{\mu_\theta}{\beta} a_\theta^4 = \mathrm{const} \equiv L
\end{equation}
is another integral of equations (\ref{eq:qpo-ar}),
(\ref{eq:qpo-atheta})  and (\ref{eq:qpo-gamma}).

\subsection{Analytical results}
Knowing two integrals of motion, we are able to find one differential
equation which governs the time-evolution of the system. 

First, the amplitudes $ a_r $ and $ a_\theta $ are not independent
because they are related by equation (\ref{eq:32_E}). To satisfy this
relation, let us define new variable $\xi(t)$ by
\begin{equation}
  \label{eq:32_xi}
  a_r^2 = \xi E,
  \quad
  a_\theta^2 = (1-\xi)\frac{E}{\nu}.
\end{equation}

\begin{figure}
\includegraphics[width=0.48\textwidth]{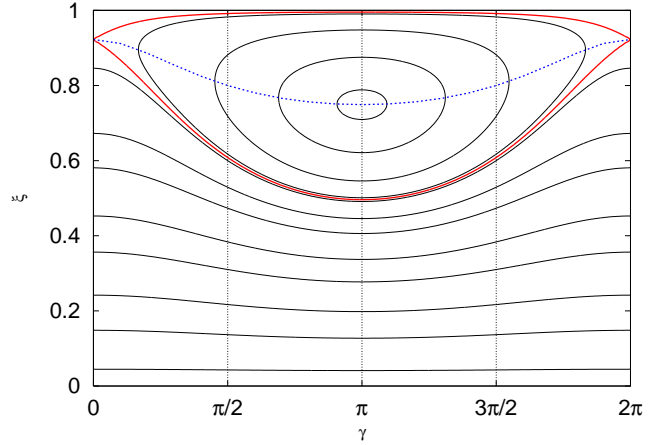}
\caption{
  Example of the $(\gamma,\xi)$-plane for the system close to the 3:2
  resonance. The  oscillations are coupled by nonlinear functions
  $f_\rho$ and $f_\theta$  [see equations (\ref{eq:res_gov_r}) and
  (\ref{eq:res_gov_theta})]. These functions give us values of  the
  constants $\alpha$, $\beta$, $\kappa_r$, $\kappa_\theta$. $\lambda_r$
  and $\lambda_\theta$. The thick solid line is the separatrix dividing the
  librating and circulating trajectories  The blue dotted line connects
  points where $\dot{\gamma}=0$. The example is for values
  $\alpha=\beta=\kappa_r=\lambda_\theta = 1$, 
  $\kappa_\theta=\lambda_\theta=2$, $\mathcal{E}=0.1$ and $\sigma =
  -0.165$.}
\label{fig:phpl}
\end{figure}

For present moment, we ignore the time dependence by  considering
projections of solutions into the $(\gamma,\xi)$-plane.  For a fixed
energy $E$ of oscillations, the system follows curves of constant  $L$.
Hence, the trajectories in the $(\gamma,\xi)$-plane are given  by
equation
\begin{equation}
  L(\gamma,\xi)=\const.
\end{equation}
An example of the phase-plane is given in Figure~\ref{fig:phpl}. There
are two types of  trajectories $[\xi(t),\gamma(t)]$: the
\imp{circulating} trajectories take the full range
$0\leq\gamma(t)\leq2\pi$ and the \imp{librating} trajectories that are
confined in the smaller range $\gamma_1\leq\gamma(t)\leq\gamma_2$. The
turning points on the librating trajectories correspond to
$\gamma=\gamma_1$ and $\gamma=\gamma_2$. This division has an
interesting consequences with respect to the observed frequencies of
resonant oscillations. According to the relation (\ref{eq:relation}) the
observed frequencies  are in exact 3:2 ratio when the system pass
through these points. On the other hand, the circulating trajectories do
not contain any turning points and the ratio of observed frequencies is
always above or bellow 3:2.

The equation describing the evolution of $\xi(t)$ can be derived in the
following way. Let us multiply equation (\ref{eq:qpo-ar}) by $ 2 a_r $
and integrate it. We obtain
\begin{equation}
  \frac{d (a_r^2)}{dt} = \frac{\alpha}{8}\omega_r a_r^3 a_\theta^2 \sin
  \gamma.
\end{equation}
Then we express $ a_r^2 $ using $ \xi $, and square it. We find
\begin{equation}
  \label{eq:32_derxi}
  \left(\frac{8 E}{\alpha \omega_r} \right)^2 \dot{\xi}^2 = \left( a_r^3
  a_\theta^2 \sin \gamma \right)^2.
\end{equation}
The right-hand side of this equation can be expressed using equation
(\ref{eq:32_L}) as
\begin{eqnarray}
  \left( a_r^3 a_\theta^2 \sin \gamma \right)^2  &=&  
  \left( a_r^3 a_\theta^2 \right)^2-
  \nonumber\\
  &\phantom{=}&\left( L  -  \frac{8\sigma}{\beta
  \omega_\theta} a_\theta^2  -  \frac{\mu_r}{\beta \nu} a_r^4  + 
  \frac{\mu_\theta}{\beta} a_\theta^4 \right)^2.
\end{eqnarray}

\begin{figure}
\includegraphics[width=0.48\textwidth]{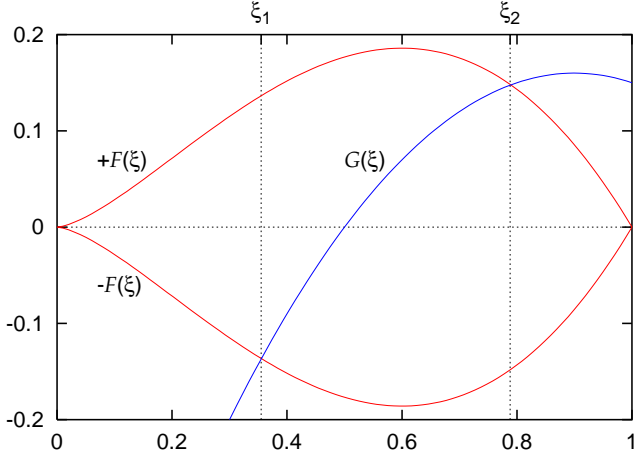}
\caption{
  The functions  $ \pm F(\xi) = \pm (1-\xi)\xi^{3/2} $ and the quadratic
  function $ G(\xi) $ whose second powers are first and second terms on
  the right-hand side of equation (\ref{eq:32_EOM}). The behavior of
  the system corresponds to $ \xi $ in the interval $ [\xi_1, \xi_2] $
  (denoted by the two dotted vertical lines) where the condition $
  |F(\xi)| \geq |G(\xi)| $ is satisfied. }
  \label{fig:32_FG}
\end{figure}

After the substitution into equation (\ref{eq:32_derxi}) and using
the relations (\ref{eq:32_xi}), we get 
\begin{eqnarray}
  \frac{1}{E^3} \left( \frac{8}{\beta\omega_\theta} \right)^2 \dot{\xi}^2 
  &=&(1-\xi)^2\xi^3  - 
  \nonumber \\
  &\phantom{=}&
  \frac{\nu^2}{E^5} \big[ L  -  \frac{8 \sigma
  E}{\beta \nu \omega_\theta} (1-\xi)  
  \nonumber \\
  &-&  \frac{\mu_r E^2}{\beta \nu}
  \xi^2  +  \frac{\mu_\theta E^2}{\beta \nu^2} (1-\xi)^2 \big]^2.
  \label{eq:32_EOM}
\end{eqnarray}
The equation of motion has a form
\begin{equation}
  \label{eq:32_EOM_form}
  \mathcal{K}^2 \dot{\xi}^2 = F^2(\xi) - G^2(\xi),
\end{equation}
where the $\mathcal{K}^2$ is a positive constant,
$F(\xi)=(1-\xi)\xi^{3/2}$ and  $G(\xi)$ is a quadratic function
coefficients of which depend on initial conditions through $E$ and $L$. 
The motion occurs only for $\xi$ that satisfy $|F(\xi)| \geq |G(\xi)|$. 
The turning points, where $ \dot{\xi} $ changes its signature, are
determined by the condition
\begin{equation}
  \label{eq:32_turn}
  |F(\xi)| = |G(\xi)|.
\end{equation}

The functions $\pm F(\xi)$ and $G(\xi)$ are plotted in Figure 
\ref{fig:32_FG}.  Generally, the function $G$ intersects
the functions $\pm F$ in two points that corresponds to $\xi(t)$
oscillating between the two bounds $\xi_1$ and $\xi_2$ given by
condition (\ref{eq:32_turn}). The radial and vertical mode of
oscillations periodically exchanges the energy. The amount of exchanged
energy is given by $ \Delta E/E = \xi_2 - \xi_1 $. For some particular 
values of $L$ and $E$ only one intersection of $\pm F$ and
$G$ exists (the function $G(\xi)$ touch one of the functions $\pm
F(\xi)$)-- the oscillations of the system correspond to the
steady-state solutions  discussed above.

\begin{figure}
  \begin{center}
    \includegraphics[width=0.48\textwidth]{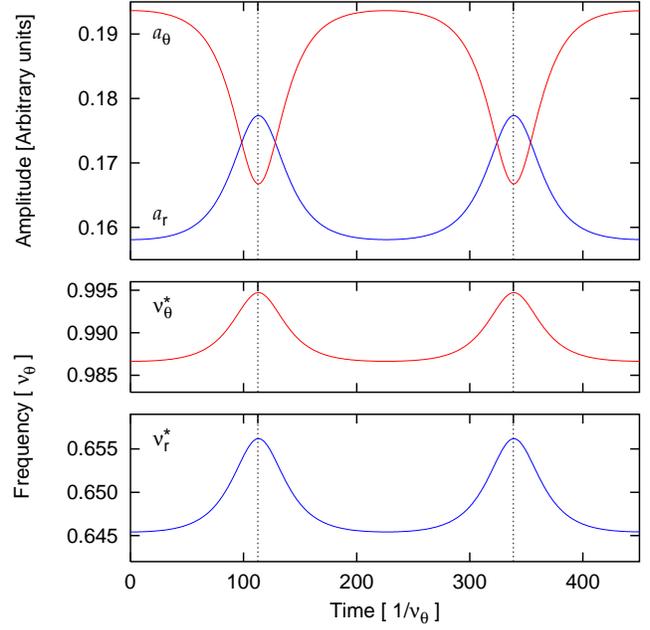}
  \end{center}
  \caption{
    Time evolution of the amplitudes (top panel) and the observed    
    frequencies, $\nu_\theta^\star = \obsom_\theta/(2\pi)$ (middle panel)
    and $\nu_r^\star = \obsom_r/(2\pi)$ (bottom panel). All quantities are
    rescaled with respect to the higher eigenfrequency $ \nu_\theta $.
    Amplitudes of the oscillations are anticorrelated because the energy is
    conserved. Observed frequencies are correlated because the system is in
    parametric resonance.}
  \label{fig:32_sol}
\end{figure}

The period of the energy exchange can be find by integration of
equation (\ref{eq:32_EOM})
\begin{equation}
  T = \frac{16}{\beta \omega_\theta} E^{-3/2} \int_{\xi_1}^{\xi_2}
  \frac{d\xi}{\sqrt{F^2(\xi) - G^2(\xi)}}.
\end{equation}
This integral can be roughly approximated as
\begin{equation}
  \label{eq:res32_T}
  T \sim \frac{16 \pi}{\beta \omega_\theta} E^{-3/2}.
\end{equation}
However, near the steady state the period becomes much longer.

The observed frequencies $\obsom_r$ and $\obsom_\theta$ are given by
relations (\ref{eq:relation}). They depend on squared amplitudes
$a_r^2$ and $a_\theta^2$. Since both $a_r^2$ and $a_\theta^2$ depend
linearly on $\xi(t)$, also observed frequencies are linear functions of
$\xi$ and are linearly correlated. The slope of this correlation
$\obsom_\theta = K \obsom_r + Q$ is independent of the energy of
oscillations and is given only by parameters of the system,
\begin{equation}
  K = \frac{\omega_\theta}{\omega_r} \frac{\lambda_r \nu -
  \lambda_\theta}{\kappa_r \nu - \kappa_\theta}.
\end{equation}
The slope of the correlation differs from 3:2, however the observed
frequencies are still close to it.

\subsection{Numerical results}

The equations (\ref{eq:qpo-ar}), (\ref{eq:qpo-atheta}) and
(\ref{eq:qpo-gamma}) were solved numerically using the fifth-order
Runge-Kutta method with an adaptive step size. One of the solutions is
shown in Figure \ref{fig:32_sol}.  The top panel of the figure shows the
time behavior of the amplitudes of the resonant oscillations. Since
energy of the system is constant, amplitudes are anticorrelated and the
two modes are continuously exchanging energy between each other.  The
middle and the bottom panels show the two observed frequencies that are
mutually correlated. They are also correlated to one of the amplitudes.
The frequency ratio varies with time and it differs from the exact 3:2
ratio, however, it always remains very close to it.  Our numerical
solution is in agreement with the general results obtained analytically
in the previous section. 

\section{Conclusions}

Although this paper was originally motivated by observations and models
connected to high-frequency QPOs,  our results are very general and can
be applied to any system with governing equations of the form
(\ref{eq:res_gov_r}) and (\ref{eq:res_gov_theta}). 

The main result of the calculations is our prediction of low-frequency
modulation of the amplitudes and frequencies of oscillations.  The
characteristic timescale of the modulation is approximately given by
equation (\ref{eq:res32_T}).

Because of the generality of our approach this fact have an interesting 
consequences in the context of QPO nature of which are unknown. Our
result can be summarized in the following way: If the two quasiperiodic
oscillations observed close to 3:2 ratio  are produced by the
autoparametric resonance the frequencies and  amplitudes of oscillations
should be periodically modulated. This modulation appears as a separate
peak at the modulation frequency and as side-bands to the main (linear)
oscillation. In a separate paper by \citet{h04} we pointed to possible
connection of this modulation with the \quot{normal branch
oscillations} (NBOs) that are often present together with QPOs. 
Specifically, we suggest that the correlation between the higher
frequency and the lower amplitude, evident in Figure \ref{fig:32_sol},
is the same  as was recently reported in Sco X-1 by \citet{ykj01}. We note
that similar behavior was recently observed also in the galactic black-hole 
candidate XTE~J1550-564 \citep{ykf02}.

\bigskip

\acknowledgements
It is a pleasure to thank Vladim{\'\i}r Karas, Marek Abramowicz, Wlodek
Klu\'zniak, Paola Rebusco, Michal Bursa and Michal Dov\v{c}iak for
helpful discussions. This work was supported by the GACR grant
205/03/0902 and GAUK grant 299/2004.



\begin{thebibliography}{}
\bibitem[Abramowicz et al., 2002] 
{abbk02} Abramowicz M.A., Bulik T., Bursa M., Klu\'{z}niak W., 2002, A\&A, 404, L21
\bibitem[Abramowicz et al., 2003]
{akklr03} Abramowicz M.A., Karas V., Klu\'{z}niak W., Lee W.H., Rebusco P., 2003, PASJ, 55, 467
\bibitem[Abramowicz \& Klu\'{z}niak, 2001]
{ak01} Abramowicz M.A., Klu\'{z}niak W., 2001, A\&A, 374, L19
\bibitem[Abramowicz \& Klu\'zniak, 2003]
{ak03} Abramowicz M.A., Klu\'{z}niak W., 2003, GReGr, 35, 69
\bibitem[Abramowicz et al., 2004]
{akst04} Abramowicz M.A., Klu\'{z}niak W., Stuchl\'{\i}k S., T\"or\"ok G., 2004, astro-ph/0401464
\bibitem[Hor\'ak et al., 2004]
{h04} Hor\'ak J., Abramowicz M.A., Karas V., Klu\'zniak W., 2004, PASJ, in press
\bibitem[Hor\'ak et al., 2005]
{hor05} Hor\'ak J., 2005, Doctoral thesis
\bibitem[Karas, 1999]
{ka99} Karas V., 1999, PASJ, 51, 317
\bibitem[Kato, 2003]
{kato03} Kato S., 2003, PASJ, 55, 801
\bibitem[Kato, 2004]
{kato04} Kato S., 2004, PASJ, 56, 559
\bibitem[Klu\'{z}niak \& Abramowicz, 2000]
{ka01} Klu\'{z}niak W., Abramowicz M.A., 2001, Acta Phys. Pol. B, B32, 3605
\bibitem[Lai, 1999]
{lai99} Lai D., 1999, ApJ, 524, 1030
\bibitem[Li \& Narayan, 2004]
{li04} Li Li-Xin, Narayan R., 2004, ApJ, 601, 414
\bibitem[Nayfeh, 1973]
{n73} Nayfeh A.H., 1973, \quot{Perturbation methods} (John Wiley \& sons, New York)
\bibitem[Nayfeh \& Mook, 1979]
{nm79} Nayfeh A.H., Mook D.T., 1979, \quot{Nonlinear oscillations} (John Wiley \& sons, New York)
\bibitem[Rebusco, 2004]
{r04} Rebusco P., 2004, PASJ, 56, 553
\bibitem[Remillard et al., 2002]
{rmmo02} Remillard R.A., Muno M.P., McClintock J.E., Orosz J.A., 2002, ApJ, 580, 1030
\bibitem[Rezzola et al., 2003]
{rezzola03} Rezzolla L., Yoshida S'i., Zanotti O., 2003, MNRAS, 344, 978
\bibitem[Schnittman \& Bertchinger, 2004]
{sb04} Schnittman J.D., Bertschinger E., 2004, ApJ, 606, 1098
\bibitem[Titarchuk, 2002]
{tit02} Titarchuk L., 2002, ApJ, 578, L71
\bibitem[Yu et al., 2001]
{ykj01} Yu W., van der Klis M., Jonker P.G., 2001, ApJ, 559, L29
\bibitem[Yu et al., 2002]
{ykf02} Yu W., van der Klis M., Fender R.P., 2002, in New Views on microquasars, eds. Ph. Durouchoux et al.
(Center for Space Physics: Kolkate), p.~72
\end{thebibliography}
\end{document}